\newcommand{\half}{\frac{1}{2}}
\newcommand{\th}{\theta}
\newcommand{\k}{\alpha}

\newcommand{\extraspace}{\addtolength{\abovedisplayskip}{2mm}
                        \addtolength{\belowdisplayskip}{2mm}
                        \addtolength{\abovedisplayshortskip}{2mm}
                        \addtolength{\belowdisplayshortskip}{2mm}}
\newcommand{\be}{\begin{equation}\extraspace}

\newcommand{\ee}{\end{equation}}
\newcommand{\bea}{\begin{eqnarray}\extraspace}
\newcommand{\beastar}{\begin{eqnarray*}\extraspace}
\newcommand{\eea}{\end{eqnarray}}
\newcommand{\eeastar}{\end{eqnarray*}}
\newcommand{\nonu}{\nonumber \\[2mm]}
\newcommand{\np}{Nucl.Phys.\ }
\newcommand{\pr}{Phys.Rev.\ }
\newcommand{\cmp}{Comm.Math.Phys.\ }
\newcommand{\pl}{Phys.Lett.\ }

\newcommand{\Rocek}{Ro\v cek}

\newcommand{\bth}{\bar{\theta}}

\newcommand{\bD}{\overline D}

\newcommand{\pa}{\partial}

\newcommand{\tz}{\frac{\theta_{12}}{z_{12}}}
\newcommand{\zot}{\frac{1}{z_{12}}}

\newcommand{\cQ}{\cal Q}
\newcommand{\cT}{\cal T}
\newcommand{\cW}{\cal W}
\newcommand{\cWc}{{\cal W}_{\frac{G}{H}}}
\newcommand{\cTc}{{\cal T}_{\frac{G}{H}}}
\newcommand{\bt}{\beta}

\newcommand{\tzb}{\frac{\bar{\theta}_{12}}{z_{12}}}
\newcommand{\tzzbb}{\frac{\theta_{12} \bar{\theta}_{12}}{z_{12}^2}}
\newcommand{\tzbb}{\frac{\theta_{12} \bar{\theta}_{12}}{z_{12}}}
\newcommand{\Ht}{\tilde{h}}

\documentstyle[12pt]{article}
\begin{document}
\thispagestyle{empty}
\begin{flushright}
JINR E2-94-? \\
hep-th/9410170\\
Oct., 1994 \\
\end{flushright}
\vspace{1cm}
\begin{center}
EXPLICIT CONSTRUCTION OF $N=2$ $W_3$ CURRENT \\
IN THE $N=2$ COSET $\frac{SU(3)}{SU(2) \times U(1)}$ MODEL\\
\vspace{1cm}
{\sc Changhyun Ahn}\footnote{email: ahn@thsun1.jinr.dubna.su}\\
\vspace{1cm}
{\it Bogoliubov  Theoretical Laboratory,\\
JINR, Dubna, Head Post Office, P.O.Box 79,\\
101 000, Moscow, Russia }\\
\vspace{3cm}
{ABSTRACT}
\end{center}
We discuss the nonlinear extension of $N=2$ superconformal algebra
by generalizing
Sugawara construction and coset construction built from $N=2$
current algebra based on Kazama-Suzuki
$N=2$ coset model $\frac{SU(3)}{SU(2) \times U(1)}$ in $N=2$
superspace. For the
generic unitary minimal series $c = 6(1-\frac{3}{k+3})$ where $k$ is
the level of $SU(3)$ supersymmetric Wess-Zumino-Witten model, this algebra
reproduces exactly $N=2$
$W_3$ algebra which has been worked out by Romans in component formalism.

\vspace{1cm}

\vfill
\setcounter{page}0
\setcounter{footnote}0
\newpage
\setcounter{equation}{0}

In recent years important progress has been made in understanding the
structure of nonlinear extension of ( super ) conformal algebra
in two dimensional
rational conformal field theory. There are three approaches \cite{BS}
which can be
used to investigate extended Virasoro symmetries, $ W $ symmetries.
One of them, third approach, is to begin with the Wess-Zumino-Witten ( WZW )
conformal field
theory and to construct extra symmetry currents from the basic fields taking
values in the underlying finite dimensional Lie algebra in that model.

In this approach,
a generalization of Sugawara construction, Casimir construction,
which includes higher spin
generator besides the stress energy tensor in terms of
currents was first presented in
\cite{BBSSa}. These analysis
for level-$1$ WZW models
provided $W$ algebras associated with the
simply laced classical Lie algebras $ADE$ in which the application of
current algebra played an important role.
Furthermore, the extension
\cite{BBSSb} to a coset \cite{coset}
construction led to study the unitary minimal models for the bosonic
$W_3$ algebra with the central charge given by
$
c_{N=0}=2 \left[1-\frac{12}{(k+3)(k+4)} \right], \;\;\; k=1,2, \ldots
$

{}From the fact that the appropriate extended algebra for a specific series
of $ A_{2}^{(1)} $ coset models of level $ (3,k) $ is an extension of $N=1$
Virasoro algebra and the first model in this series, with level $
(1,k) $ has bosonic $W_{n}$ symmetry, it has been developed further
in the case of $N=1$ $W_{3}$ algebra \cite{ASS} that for
$
c_{N=1}=4 \left[1-\frac{18}{(k+3)(k+6)} \right], \;\;\; k=1,2, \ldots ,
$
the complete set of supercurrents of it should contain eight supercurrents.
Of course the only five
supercurrents were constructed by computing the operator product
expansions ( OPE's )
explicitly and came up with final analysis using so-called 'character
technique'.

It is a natural question to ask for $N=2$ $W_3$ case.
It was shown \cite{KS} that a large class of unitary conformal
field theories with
$N=2$ superconformal symmetry can be realized as coset model $\frac{G}{H}$
called
hermitian symmetric spaces. These $N=2$ model based on $\frac{
SU(3)}{SU(2) \times U(1)} $ has an $N=2$ $ W_3 $ algebra \cite{Ro} as its
chiral algebra by duality symmetry of compact Kazama-Suzuki models
and has a central charge,
\be
c_{N=2}=6(1-\frac{3}{k+3}), \;\;\; k=1,2, \ldots
\label{eq:central}
\ee

It is very instructive to consider these minimal models in the context of
$N=2$ supersymmetric extension of the affine Lie algebra that was
given \cite{HS}
in $N=2$ superspace in terms of supercurrents satisfying nonlinear
constraints. We would like to apply this supercurrent algebra to
supersymmetric WZW conformal
field theory and understand how higher spin $2$ supercurrent appears
in $SU(3)$  $N=2$ affine Lie algebra.

In this paper, we make an attempt to identify the independent
generating supercurrents in terms of the basic superfields satisfying
the nonlinear constraints through a generalized Sugawara construction and
coset construction,
as we will see below, in $N=2$ $CP_2$ model
in the above series (\ref{eq:central}).

Let us first consider a few things about the on-shell current algebra
\cite{HS,RASS}
in $N=2$ superspace for the supersymmetric WZW model, with level $k$,
on a group $G=SU(3)$.
We choose a complex basis for the Lie algebra, labelled by $a$, $\bar{a}$,
$a=1,2,\ldots$, $\frac{1}{2} \dim \, G (=4 \;\mbox{in the adjoint
representation of $G$}) $, in which the complex
structure related to the second supersymmetry has eigenvalue
$+i$ on the hermitian generators $T_a$ and $-i$ on the conjugated generators
$T_{\bar{a}} (=T_a^{\dagger})$.
In this complex basis, they satisfy the following relations:
$
[T_a,T_b]={f_{ab}}^c T_c,\; [T_a,T_{\bar{b}}]={f_{a\bar{b}}}^c T_c+
{f_{a\bar{b}}}^{\bar{c}} T_{\bar{c}},\;
Tr(T_a T_b)=0,\;
Tr(T_a T_{\bar{b}})=\delta_{a\bar{b}}
$ where $f$'s are the structure constants.

Then the $N=2$ currents
${\cQ}^a$ and ${\cQ}^{\bar{a}}$ can be characterized by the {\it nonlinear}
constraints. We will only discuss the {\it "chiral"} currents in the
sense that they are annihilated by $D_-$ and $\bD_-$. $ ( D_- {\cQ}^a=
\bD_- {\cQ}^a=D_- {\cQ}^{\bar{a}}=\bD_- {\cQ}^{\bar{a}}=0. )$ For brevity
we will write $D$ for $D_+$ and $\bD$ for $\bD_+$.
\be
D {\cQ}^a=-\frac{1}{2(k+3)} {f^a}_{bc} {\cQ}^b {\cQ}^c, \quad
\bD {\cQ}^{\bar{a}}=-\frac{1}{2(k+3)} {f^{\bar{a}}}_{\bar{b}
     \bar{c}} {\cQ}^{\bar{b}} {\cQ}^{\bar{c}} \,.
\label{eq:cons}
\ee
Here, we work with complex spinor covariant derivatives
\bea
D=\frac{\partial}{\partial \theta}-\frac{1}{2} \bar {\theta}
\partial,\;\;\;\;\; \bD =\frac{\partial}{\partial \bar{\theta}}-\frac{1}{2}
\theta \partial \nonu
\eea
satisfying the algebra
\bea
\{ D, \bD\}=-\partial\;(=-\partial_{z}),
\eea
all other anticommutators vanish\footnote{Notice that
we take different conventions
from those of ref. \cite{RASS}}.
The dual Coxeter number of $SU(3)$, $\Ht$, is replaced by $3$ in the eq.
(\ref{eq:cons}) and the $f$'s
are antisymmetric in lower two indices\footnote{ We choose the structure
constants as follows:
$
{f_{14}}^3={f_{3\bar{4}}}^1={f_{4\bar{3}}}^{\bar{1}}=-1,\;
{f_{1\bar{1}}}^{\bar{2}}={f_{1\bar{2}}}^1=-i,\;
{f_{23}}^3={f_{3\bar{3}}}^{\bar{2}}={f_{4\bar{4}}}^2=-\half (\sqrt{3}+i)
, \;
{f_{24}}^4={f_{3\bar{3}}}^2={f_{4\bar{4}}}^{\bar{2}}=-\half (\sqrt{3}-i).
$
}. Any indices can be raised and lowered using $\delta^{a\bar{b}} \mbox{and
} \delta_{a\bar{b}}$.
The fundamental OPE's
of these superfields are
\bea
 {\cQ}^a (Z_{1}) {\cQ}^b (Z_{2})   & = & -\tzb {f^{ab}}_{c} {\cQ}^c
-\tzbb \frac{1}{k+3} {f^{a}}_{ec} {f^{be}}_{d}
 {\cQ}^c {\cQ}^d
\nonu
 {\cQ}^{\bar{a}} (Z_{1}) {\cQ}^{\bar{b}} (Z_{2})  & = &
 -\tz {f^{\bar{a} \bar{b}}}_{\bar{c}} {\cQ}^{\bar{c}}
 +\tzbb \frac{1}{k+3} {f^{\bar{a}}}_{\bar{e} \bar{c}}
 {f^{\bar{b} \bar{e}}}_{\bar{d}} {\cQ}^{\bar{c}} {\cQ}^{\bar{d}}
\nonu
 {\cQ}^{a} (Z_{1}) {\cQ}^{\bar{b}} (Z_{2})  & = &
 \tzzbb \frac{1}{2} \left[(k+3) \delta^
{a \bar{b}}  + {f^a}_{cd} f^{\bar{b} cd} \right]
  -\frac{1}{z_{12}} (k+3) \delta^{a \bar{b}} \nonu
 & & - \tz {f^{a \bar{b}}}_c {\cQ}^c  - \tzb
 {f^{a \bar{b}}}_{\bar{c}} {\cQ}^{\bar{c}}  \nonu
 & & - \tzbb \left[ {f^{a \bar{b}}}_c \bar{D} {\cQ}^c
 + \frac{1}{k+3} {f^{a \bar{c}}}_{d} {f^{\bar{b}}}_{\bar{c} \bar{e}}
 {\cQ}^d {\cQ}^{\bar{e}}  \right],
\label{eq:qq}
\eea
where
\be
\th_{12}=\th_{1}-\th_{2}, \; \bth_{12}=\bth_{1}-\bth_{2}, \;
z_{12}=z_{1}-z_{2}+\frac{1}{2}(\th_{1} \bth_{2} + \bth_{1} \th_{2})
\ee
and all the superfields in the right hand side are evaluated at $Z_2$.
As we will see in later, these OPE's can be used as the basis for a
generalized Sugawara construction and coset construction. It can be checked
 that the Jacobi identities of this
algebra are satisfied only if we have the above nonlinear constraints
(\ref{eq:cons}). The quadratic terms in (\ref{eq:qq}) are essential for
this consistency. Of
course, the familiar reduction to $N=1$ unconstrained superfield description
of (\ref{eq:qq}) has been discussed
already in \cite{RASS}.

Let us now focus on the $N=2$ superconformal algebra.
The appropriate generalization to $N=2$ superspace of the well-known
Sugawara construction gives the following formula for the $N=2$
stress tensor in terms of the supercurrents ${\cQ}^a$ and
${\cQ}^{\bar{a}}$ (\cite{HS}),
\bea
{{\cal T}_G} &=& -\frac{1}{k+3} \delta_{a \bar{b}} {\cQ}^a {\cQ}^{\bar{b}}
  +\frac{1}{k+3} ( \delta_{b \bar{c}} f_{\bar{a}}{}^{b
  \bar{c}} D {\cQ}^{\bar{a}}
  + \delta_{b \bar{c}} f_a{}^{b \bar{c}} \bD {\cQ}^a)\nonu
           &=&-\frac{1}{k+3} \left[ {\cQ}^1 {\cQ}^{\bar{1}}+
{\cQ}^2 {\cQ}^{\bar
{2}}+{\cQ}^3 {\cQ}^{\bar{3}}+{\cQ}^4 {\cQ}^{\bar{4}}
-(\sqrt{3}-i) D {\cQ}^{\bar{2}}-(\sqrt{3}+i) \bD {\cQ}^2 \right].
\eea
It satisfies the OPE according to (\ref{eq:qq}),
\be
{{\cT}_G} (Z_{1}) {{\cT}}_G (Z_{2})=  \frac{1}{z^{2}_{12}} \frac{c_G}{3}
  + \left[ \tzzbb -\tz D+\tzb \bD +\tzbb \partial_2 \right]
{{\cT}}_G  \,.
\label{eq:tgtg}
\ee
The central charge is
\be
c_G = \frac{3 \,\dim G}{2} \left[1-\frac{2 \Ht}{3 (k+\Ht)} \right]
=12\frac{k+1}{k+3}.
\ee
The $N=2$ superfield ${\cal T}_G$ has
as its component fields the bosonic stress tensor $T$ of spin $2$, two
supercurrents
$G^{+}$ and $G^{-}$ of spins $3/2$ and the $U(1)$ current $J$ of spin $1$,
which together form
the familiar $N=2$ current algebra.

Let $H=SU(2) \times U(1)$ be a subgroup of $G(=SU(3))$ and denote the indices
corresponding to $H$ and the coset $\frac{G}{H}=\frac{SU(3)}{SU(2) \times
U(1)}$ by $1, {\bar{1}}, 2, {\bar{2}}$ and $3, {\bar{3}}, 4, {\bar{4}}$,
respectively. Then $N=2$ currents can be divided into two
sets. We may construct another $N=2$ superconformal algebra,
\bea
{{\cal T}_H} =-\frac{1}{k+3} \left[ {\cQ}^1 {\cQ}^{\bar{1}}+
{\cQ}^2 {\cQ}^{\bar
{2}}+i D {\cQ}^{\bar{2}}-i \bD {\cQ}^2 \right].
\label{eq:th}
\eea
It satisfies the OPE
\be
{{\cT}_H} (Z_{1}) {{\cT}}_H (Z_{2})=  \frac{1}{z^{2}_{12}} \frac{c_H}{3}
  + \left[ \tzzbb -\tz D+\tzb \bD +\tzbb \partial_2 \right]
{{\cT}}_H  \,.
\label{eq:thth}
\ee
The total central charge is the sum of contributions $
\frac{3 \,\dim H}{2} \left[1-\frac{2 \Ht}{3 (k+\Ht)} \right] $ for each simple
factor of $H$.
That is,
\be
c_H=\frac{9}{2} \left[1-\frac{4}{3(k+2)} \right]+\frac{3}{2}=
6\frac{k+2}{k+3}.
\ee
It is easy to see that a realization of 'large' $N=4$ superconformal
algebra \cite{n4} with $k_{+}=k+2,\; k_{-}=1$ where the two affine
$SU(2)$ subalgebras have level $k_{+}$ and $k_{-}$ respectively, can be
generated by $\{ {\cQ}^{2, \bar{2}},\;\; {\cQ}^{1},\;\;{\cQ}^{\bar{1}},\;\;
{\cT}_{H} \}$ as shown in \cite{RASS}.

We take a closer look at the supersymmetric coset models based on
$N=2$ $CP_2$ model with level $k$ of $SU(3)$. Let us define
\bea
{\cT}_{\frac{G}{H}}={\cT}_G-{\cT}_H
=-\frac{1}{k+3} \left[ {\cQ}^3 {\cQ}^{\bar{3}}+
{\cQ}^4 {\cQ}^{\bar
{4}}-\sqrt{3} D {\cQ}^{\bar{2}}-\sqrt{3} \bD {\cQ}^2 \right].
\label{eq:tc}
\eea
Using the eqs. (\ref{eq:qq}), we can conclude that
\be
{\cT}_{\frac{G}{H}} (Z_1) {\cQ}^{1} (Z_2)=
{\cTc} (Z_1) {\cQ}^{\bar{1}} (Z_2)={\cTc} (Z_1) {\cQ}^{2} (Z_2)=
{\cTc} (Z_1) {\cQ}^{\bar{2}} (Z_2)=0
\label{eq:reg1}
\ee
and then
 $ {\cT}_{\frac{G}{H}} (Z_1) {\cT}_{H} (Z_2)=0$.
This decomposition implies from (\ref{eq:tgtg}) and (\ref{eq:thth})
that ${\cT}_{\frac{G}{H}}$ satisfies the OPE
\be
{\cT}_{\frac{G}{H}} (Z_{1}) {{\cT}}_{\frac{G}{H}} (Z_{2})=
\frac{1}{z^{2}_{12}}
\frac{c_{\frac{G}{H}}}{3}
  + \left[ \tzzbb -\tz D+\tzb \bD +\tzbb \partial_2 \right]
{\cT}_{\frac{G}{H}},
\label{eq:tctc}
\ee
where
\be
c_{\frac{G}{H}}=c_G-c_H=6\frac{k}{k+3}
\ee
which is the same as (\ref{eq:central}) ( as a function of $k$ ).
The super currents $ \{ {\cQ}^{3}, {\cQ}^{\bar{3}}, {\cQ}^{4},
{\cQ}^{\bar{4}} \}$
are not $N=2$ primary fields because of nonlinear constraints among them.
For example,
\bea
& & {\cT}_{\frac{G}{H}} (Z_{1}) {\cQ}^{3} (Z_{2}) =\nonu
& & \tzzbb
\frac{k}{2(3 + k)} {\cQ}^{3}+
   \zot \frac{k}{(3 + k)} {\cQ}^{3} + \tzb \left[ \bD {\cQ}^{3} +
    \frac{(i - \sqrt{3})}{2(3 + k)} {\cQ}^{\bar{2}} {\cQ}^{3} \right] \nonu
   & & +\tzbb \left[ \frac{1}{(3 + k)^2} {\cQ}^{1} {\cQ}^{3} {\cQ}^{
\bar{1}} +
     \frac{i}{(3 + k)^2} {\cQ}^{\bar{2}} {\cQ}^{1} {\cQ}^{4} \right.\nonu
& &\left.+
     \frac{1}{(3 + k)} \bD {\cQ}^{1} {\cQ}^{4} +
    \frac{(i+\sqrt{3})}{2(3 + k)} \bD {\cQ}^{2} {\cQ}^{3} +
    \frac{(-i + \sqrt{3})}{2(3+k)} D {\cQ}^{\bar{2}} {\cQ}^{3}
      +  \partial {\cQ}^{3} \right]. \nonu
\eea
And the $U(1)$ charge is the fractional function of $k$, $k/(3+k)$.
However we do not write dowm explicitly for other OPE's $\cTc$ with
$\{ {\cQ}^{\bar{3}}, {\cQ}^{4}, {\cQ}^{\bar{4}} \} $ here, they are
necessary for (\ref{eq:tw}).

In order to extend the coset construction to the higer spin current we
proceed as follows \cite{BBSSb}.
Sugawara's expression for the higher spin current can be obtained as
composite operators of currents contracted with $ \delta \;
\mbox{and} \; f $
tensors. Notice that each term should have the correct $ U(1) $ charge
conservation, that is, $\cWc$ has vanishing $U(1)$ charge.
The dimension $2$ coset field ${\cal W}_{
\frac{G}{H}}$ is uniquely fixed by the requirements that it should be a
superprimary field of dimension $2$ with respect to $ \cTc $
and the OPE's with
$ \{ {\cQ}^{1}, {\cQ}^{\bar{1}}, {\cQ}^{2}, {\cQ}^{\bar{2}} \}$ are regular
up to overall constant $A(k)$:
\be
{\cW}_{\frac{G}{H}} (Z_1) {\cQ}^{1} (Z_2)={\cWc} (Z_1) {\cQ}^{\bar{1}}
(Z_2)={\cWc} (Z_1) {\cQ}^{2} (Z_2)={\cWc} (Z_1) {\cQ}^{\bar{2}} (Z_2)=0,
\label{eq:reg}
\ee
\be
{\cT}_{\frac{G}{H}} (Z_{1}) {{\cW}}_{\frac{G}{H}} (Z_{2})=
   \left[ \tzzbb 2 -\tz D+\tzb \bD +\tzbb \partial_2 \right]
{\cW}_{\frac{G}{H}}.
\label{eq:tw}
\ee

Thus
\bea
{\cal W}_{\frac{G}{H}} &=&
A(k)(3+k) \left[ \frac{k(3 + k)^2(5 + k)}{2(3 - 5k)} [D,\bD]
{\cT}_{\frac{G}{H}} +
  \frac{(3 + k)^2}{2} [D,\bD] {\cT}_{H}- 2{\cQ}^2
{\cQ}^{\bar{2}} {\cQ}^4 {\cQ}^{\bar{4}} \right. \nonu
& &+
  (i + \sqrt{3}) {\cQ}^2 {\cQ}^{\bar{2}} D {\cQ}^{\bar{2}} -
  i{\cQ}^2 {\cQ}^3 {\cQ}^{\bar{1}} {\cQ}^{\bar{4}} +
  (-i + \sqrt{3}) {\cQ}^2 \bD {\cQ}^2 {\cQ}^{\bar{2}}\nonu
& & +
  (3 + k)  {\cQ}^2 \partial {\cQ}^{\bar{2}} +
  i {\cQ}^{\bar{2}} {\cQ}^1 {\cQ}^4 {\cQ}^{\bar{3}} -
  (3 + k)(5 + k) {\cQ}^4 \partial {\cQ}^{\bar{4}} \nonu
& &-
  (3 + k)(5 + k) {\cT}_{\frac{G}{H}} {\cQ}^4 {\cQ}^{\bar{4}}+
  \frac{3k(3 + k)^2(5 + k)}{2(3 - 5k)} {\cT}_{\frac{G}{H}} {\cT}_{\frac{G}{
H}}-
  (3 + k)^2 {\cT}_{\frac{G}{H}} {\cT}_{H} \nonu
& &+ \frac{(3 + k)
     (7 - 3i\sqrt{3} + k - i\sqrt{3}k)}{
   (-i + \sqrt{3})} {\cT}_{\frac{G}{H}} \bD {\cQ}^2 +
\frac{2(3 + k)(4 - i\sqrt{3} + k)
     }{(-i + \sqrt{3})} {\cT}_{\frac{G}{H}} D {\cQ}^{\bar{2}} \nonu
& &-
  (3 + k) {\cT}_{\frac{G}{H}} {\cQ}^2 {\cQ}^{\bar{2}}
-
2(3 + k) {\cT}_{H} {\cQ}^4 {\cQ}^{\bar{4}} +
  \frac{(3 + k)^2}{2} {\cT}_H {\cT}_H \nonu
& & + (-i + \sqrt{3})(3 + k) {\cT}_{H} \bD {\cQ}^2
   +
  (i + \sqrt{3})(3 + k) {\cT}_{H} D {\cQ}^{\bar{2}} +
  (3 + k) {\cT}_{H} {\cQ}^2 {\cQ}^{\bar{2}} \nonu
& &+
(3 + k) \bD {\cQ}^{1} {\cQ}^4 {\cQ}^{\bar{3}}
+
  (-i + 5\sqrt{3} - ik +\sqrt{3}k) \bD {\cQ}^2 {\cQ}^4 {\cQ}^{
\bar{4}} \nonu
& & -
  \frac{(-5i + 3\sqrt{3} - 2ik)}{(-i+\sqrt{3})}
 \bD {\cQ}^2 \bD {\cQ}^2
    - (5 + k) \bD {\cQ}^2 D {\cQ}^{\bar{2}} \nonu
& &+
  (3 + k) D {\cQ}^{\bar{1}} {\cQ}^3 {\cQ}^{\bar{4}} +
  (i + 5\sqrt{3} + ik + \sqrt{3}k) D {\cQ}^{\bar{2}} {\cQ}^4
{\cQ}^{\bar{4}} \nonu
& &-
  \frac{(-2i + 4\sqrt{3} + ik + \sqrt{3}k)}{(-i+\sqrt{3})}
     D {\cQ}^{\bar{2}} D {\cQ}^{\bar{2}} -
  (3 + k) \partial {\cQ}^{2} {\cQ}^{\bar{2}} \nonu
& &\left.+
  (3 + k)(5 + k) \partial {\cQ}^4 {\cQ}^{\bar{4}} \right].
\label{eq:qres}
\eea
In the above expression, we used the fact that the normal ordering of
${\cQ}^{3} {\cQ}^{\bar{3}} $ can be written as $-(k+3) \cTc $ besides other
terms, seen by (\ref{eq:tc}) and similarly ( $ {\cQ}^{1} {\cQ}^{\bar{1}} $
as $-(k+3) {\cT}_{H} $ ) because in the OPE $\cWc (Z_1) \cWc (Z_2)$, $
\cTc (Z_1) \cWc (Z_2) $ transforms as (\ref{eq:tw}) and ${\cT}_H (Z_1)
\cWc (Z_2)=0 $ which make the calculations easier. The coefficients of
composite operators having $1, \bar{1}, 2, \bar{2} $ indices in
(\ref{eq:qres})
can not fixed as (\ref{eq:tw}) but as (\ref{eq:reg}) because $\cTc$
commutes with $ \{ {\cQ}^{1}, {\cQ}^{\bar{1}}, {\cQ}^{2},
{\cQ}^{\bar{2}} \} $ as
(\ref{eq:reg1}). All the $ D {\cQ}^{a}, \bD {\cQ}^{\bar{a}} $'s are written
as the right hand side of (\ref{eq:cons}).
By requiring the fourth order pole of $\cWc (Z_1) \cWc (Z_2) $ should be
$c_{\frac{G}{H}}/2$, the normalization factor $A(k)$ can be fixed as follows:
\be
A(k)=\sqrt{\frac{3(-3+5k)}{(-1+k)(k+3)^6(k+5)(2k+3)}}.
\ee
After a tedious calculation by writing $\cWc (Z_1) $ as in (\ref{eq:qres})
and computing the $29$th OPE's with $\cWc (Z_2) $,
we arrive at the final result,
\bea
& & \cWc (Z_1) \cWc (Z_2) = \nonu
& & \frac{1}{z_{12}^4} \frac{3k}{(3 + k)}
 + \frac{\th_{12} {\bar{\theta}}_{12}}{z_{12}^4} 3 \cTc
  + \frac{{\bar{\theta}}_{12}}{z_{12}^3} 3 \bD \cTc -
\frac{\th_{12}}{z_{12}^3}
3 D \cTc
 + \frac{\th_{12} {\bar{\theta}}_{12}}{z_{12}^3} 3 \partial \cTc \nonu
& &   +\frac{1}{z_{12}^2} \left[ 2 \k \cWc + \frac{6k}{(3 - 5k)}
[ D, \bD ] \cTc+
    \frac{3(3 + k)}{(3 - 5k)} \cTc \cTc \right] \nonu
& & +\frac{{\bar{\theta}}_{12}}{z_{12}^2} \left[ \k \bD \cWc +
    \frac{9(-1 + k)}{(-3 + 5k)} \partial \bD \cTc +
    \frac{3(3 + k)}{(3 - 5k)} \cTc \bD \cTc \right] \nonu
& & +\frac{\th_{12}}{z_{12}^2} \left[ \k D \cWc +
    \frac{9(1 - k)}{(-3 + 5k)} \pa D \cTc +
    \frac{3(3 + k)}{(3 - 5k)} \cTc D \cTc \right] \nonu
& & +\tzzbb \left[
    \frac{\k (12 + k)}{6(-2 + k)}  [D, \bD] \cWc +
    \frac{7\k (3 + k)}{3(-2 + k)}  \cTc \cWc \right.\nonu
& & +
   \frac{9}{2} \bt (-2 + k)k(3 + k) \partial [D, \bD] \cTc
     +
    \frac{3}{2} \bt (3 + k)(6 + k + 5k^2) \cTc [D, \bD] \cTc
    \nonu
& &      +\frac{3}{2} \bt (3 + k)^2(1 + 3k) \cTc \cTc \cTc
      +
   18 \bt (-2 + k)k(3 + k) \bD \cTc D \cTc \nonu
 & &  \left.   -
    \frac{3}{2} \bt (9 - 6k - 3k^2 + 8k^3) \partial^2 \cTc
      \right] \nonu
& &+\zot \left[
   \frac{3k}{(3 - 5k)} \partial [D, \bD] \cTc+
    \frac{3(3 + k)}{(3 - 5k)} \partial \cTc \cTc+
\k \partial \cWc \right] \nonu
 & & +\tzb \left[  \frac{\k (-3 + 5k)}{
     3(2 - k)(1 + k)} \pa \bD \cWc-\frac{3}{4} \bt k(9 + 12k + 11k^2)
        \pa^2 \bD \cTc \right. \nonu
 & &     +
    \frac{\k (3 + k)(5 + k)}{
     3(-2 + k)(1 + k)} \cTc \bD \cWc
+
    \frac{3}{2} \bt (3 + k)^2(1 + 3k) \cTc \cTc \bD \cTc  \nonu
 & &     +
    \frac{\k (3 + k)(-3 + 5k)}{
     3(-2 + k)(1 + k)} \bD \cTc \cWc +
    \frac{3k(3 + k)}{
     2(1 - k)(3 + 2k)} \bD \cTc [D, \bD] \cTc \nonu
& & \left. +
    9 \bt (3 + k)(-1 + 2k + k^2)
       \pa \bD \cTc \cTc
      +
    \frac{3(3 + k)}{
     2(-1 + k)(3 + 2k)} \pa \cTc \bD \cTc \right] \nonu
& &+ \tz \left[
    \frac{\k (-3 + 5k)}{
     3(2 - k)(1 + k)} \pa D \cWc+  \frac{9}{2} \bt k(3 + k^2)
       \pa^2 D \cTc \right.
   \nonu
& &  +
    \frac{\k (3 + k)(5 + k)}{3(2 - k)(1 + k)} \cTc D \cWc-
    \frac{3}{2} \bt (3 + k)^2(1 + 3k) \cTc \cTc D \cTc \nonu
& & +
    \frac{3k(3 + k)}{2(-1 + k)(3 + 2k)} [D, \bD]
\cTc D \cTc +
    \frac{\k (3 + k)(-3 + 5k)}{
     3(2 - k)(1 + k)} D \cTc \cWc \nonu
& & \left. + 9 \bt (3 + k)(-1 + 2k + k^2)
       \pa D \cTc \cTc +
    \frac{3(3 + k)}{
     2(-1 + k)(3 + 2k)} \pa \cTc D \cTc \right] \nonu
& &+ \tzbb \left[
    \frac{ \k k(5 + k)}{
     3(-2 + k)(1 + k)} \pa [D, \bD] \cWc +
    \frac{2 \k (3 + k)(3 + 2k)}{
     3(-2 + k)(1 + k)} \cTc \pa \cWc \right. \nonu
& &+
    \frac{2 \k (3 + k)}{3(1 + k)} \bD \cTc D \cWc+
    12 \bt (-2 + k)k(3 + k)
        \pa \bD \cTc D \cTc \nonu
& & +
    3 \bt (3 + k)(3 + k^2)
        \pa [D, \bD] \cTc \cTc +
    \frac{2\k (3 + k)}{3(1 + k)} D \cTc \bD \cWc \nonu
& & -
    12 \bt (-2 + k)k(3 + k)
       \pa D \cTc \bD \cTc      +
    \frac{2\k(3 + k)(1 + 3k)}{
     3(-2 + k)(1 + k)} \pa \cTc \cWc \nonu
& &+
    3\bt k(3 + k)(1 + 3k)
     \pa \cTc  [D, \bD] \cTc   +
    3\bt (3 + k)^2(1 + 3k)
      \pa \cTc \cTc \cTc \nonu
& & \left. -
    \frac{3}{2} \bt (-3 - 4k - 3k^2 + 2k^3) \pa^3 \cTc
      \right]
\label{eq:wcwc}
\eea
where
\be
 \k= \frac{3\sqrt{3}(2-k)(1+k)}{
\sqrt{(-1+k)(5+k)(3+2k)(-3+5k)}}, \;\;\;
\bt=\frac{1}{(3-5k)(-1+k)(3+2k)}
\ee
Therefore the algebra, (\ref{eq:tctc}), (\ref{eq:tw}), and (\ref{eq:wcwc})
coincide with exactly those explained in \cite{Ro,ahn} with all
superfields replaced by their coset analogues. From the explicit OPE's
about $N=2$ current satisfying nonlinear constraints
in $SU(3)$ supersymmetric WZW model, we have determined spin $2$
supermultiplet by exploiting a generalized Sugawara construction and coset
construction in $N=2$ superspace. For the central charge less than six of
any unitary representation of this
algebra consisting of $ \{ \cTc, \cWc \} $, we have
found the full algebraic structure of it.
We realize that
our findings from the viewpoint of supersymmetric WZW model
in third approach mentioned in the introduction
reproduce those \cite{Ro,ahn} in the first approach.
An interesting aspect to study in
the future direction
is how to realize for the noncompact coset $SU(2,1)/U(2)$ model which
has the central charge $c=6(1+\frac{3}{k+2}) \;\; k=1,2, \cdots $
that contains $c=9$.

It would be very interesting to understand how our explicit form of $N=2$
$W_3$ symmetry current enables us to a basis for the Landau-Ginzberg
polynomials \cite{LS} and how to generalize our results for $CP_3$
Kazama-Suzuki model \cite{bw} in which they constructed the nonlinear
extension of $N=2$ superconformal algebra by two superprimary fields
of spin $2, 3$ with zero $U(1)$ charge and even
$CP_n$ case. For $CP_3$, we have odd dimensional groups $G=SU(4)$ and $H=
SU(3) \times U(1) $ then maybe we can construct on the even dimensional
groups $G=SU(4) \times U(1)$ and $H=SU(3) \times U(1) \times U(1)$ by
following the same procedure.
{}From the viewpoint of second approach in \cite{BS} it was
worked out that $N=2$ coset model
$CP_n=\frac{SU(n+1)}{SU(n)\times
U(1)}$ can be obtained by the quantum hamiltonian reduction of the affine
Lie superalgebra ${A(n,n-1)}^{(1)}$ and the chiral algebra of this model
is $N=2$
$W$- algebra via the quantum super Miura transformation \cite{Ito,NY,Ito1}.
It is also open problem to study how our coset $W$ algebras are
related to quantum Drinfeld-Sokolov $W$ algebras in the spirit of
\cite{Ito,NY,Ito1,IK}.

\vspace{15mm}

I would like to thank ITP at Stony Brook where the early stage of this work
was done, E. Ivanov, S. Krivonos and A. Sorin for carefully
reading the manuscript,
and K. Thielemans and S. Krivonos for their package SOPEN2defs.


\begin{thebibliography}{99}

\bibitem{BS}
 P. Bouwknegt and K. Schoutens, W-Symmetries in Conformal Field
 Theory, Phys. Reps. {\bf 223} (1993) 183

\bibitem{BBSSa}
 F.A. Bais, P. Bouwknegt, K. Schoutens and M. Surridge, Extensions
 of the Virasoro Algebra Constructed from Kac-Moody Algebras Using
 Higher Order Casimir Invariants, \np  {\bf B304} (1988) 348.

\bibitem{BBSSb}
 F.A. Bais, P. Bouwknegt, K. Schoutens and M. Surridge,
 Coset Construction for Extended
 Virasoro Algebras, \np  {\bf B304} (1988) 371.

\bibitem{coset}
 K. Bardakci and M.B. Halpern, New Dual Quark Models, \pr {\bf D3} (1971)
 2493; M.B. Halpern, The Two Faces of a Dual Pion-Quark Model, \pr
 {\bf D4} (1971) 2398; P. Goddard and D. Olive, Virasoro Algebras and
 Coset Space Models, \pl {\bf 152B} (1985) 88; Unitary Representations of
 the Virasoro and Super-Virasoro Algebra, \cmp {\bf 103} (1986) 105.

\bibitem{ASS}
 C. Ahn, K. Schoutens
 and A. Sevrin, The Full Structure of the Super $W_{3} $ Algebra,
 Int.J.Mod.Phys. {\bf A6} (1991) 3467.

\bibitem{KS}
 Y. Kazama and H. Suzuki, New $N=2$ Superconformal Field Theories and
 Superstring Compactification, \np {\bf 321B} (1989) 232; Characterization
 of $N=2$ Superconformal Models Generated by the Coset Space Method, \pl
 {\bf 216B} (1989) 112

\bibitem{Ro}
 L.J. Romans,
  The $N=2$ Super-$W_{3}$ Algebra, \np {\bf B369 } (1992) 403.

\bibitem{HS}
 C.M. Hull and B. Spence, $N = 2$ Current Algebra and Coset Models,
 \pl {\bf 241B} (1990) 357.

\bibitem{RASS}
 M. \Rocek, C. Ahn, K. Schoutens and A. Sevrin, Superspace WZW
 Models and Black Holes, in Workshop on Superstrings
 and Related Topics, Trieste, Aug. 1991., IASSNS-HEP-91/69,
 ITP-SB-91-49, LBL-31325, UCB-PTH-91/50.

\bibitem{n4}
 M. Ademollo {\it et al.}, Supersymmetric Strings and Colour
 Confinement, \pl {\bf 62B} (1976) 105;
 K. Schoutens, $ O (N) $-Extended Superconformal Field Theory in
 Superspace, \np {\bf B295} (1988) 634;
 K. Schoutens, A Non-Linear Representation of the $d = 2 \; so (4)$
 -Extended Superconformal Algebra, \pl {\bf 194B} (1987) 75;
 A. Sevrin, W. Troost and A. Van Proeyen, Superconformal Algebras in
 Two Dimensions with $ N = 4$, \pl {\bf 208B} (1988) 447;
 E. Ivanov, S. Krivonos and V. Leviant, Quantum $N=3,4 $ Superconformal
 WZW Sigma Models, \pl {\bf B215} (1989) 689, {\bf B221} (1989) 432E.

\bibitem{ahn}
 C. Ahn, Free Superfield Realization of $N=2$ Quantum
 Super $W_{3}$ Algebra, Mod. Phys. Lett. {\bf A9} (1994) 271.

\bibitem{LS}
 W. Lerche and A. Sevrin, On the Landau-Ginzburg Realization of Topological
 Gravities, CERN-TH.7210/94, hep-th/9403183.

\bibitem{bw}
 R. Blumenhagen and A. Wi$\beta$kirchen, Extension of the $N=2$ Virasoro
 algebra by two primary fields of dimension $2$ and $3$, BONN-TH-94-12,
 hep-th/9408082.

\bibitem{Ito}
 K. Ito, Quantum Hamiltonian Reduction and $N=2$ Coset Models, \pl
{\bf 259B} (1991) 73 ; $N=2$ Superconformal $CP_n$ Model, \np {\bf B370}
 (1992) 123

\bibitem{NY}
 D. Nemeschansky and S. Yankielowicz, $N=2$ W-Algebra, Kazama-Suzuki Models
 and Drinfeld-Sokolov Reduction, USC-91/005.

\bibitem{Ito1}
 K. Ito, Free Field Realization of $N=2$ Super $W_3$ Algebra,
 \pl {\bf B304} (1993) 271

\bibitem{IK}
 K. Ito and H. Kanno, Lie Superalgebra and Extended Topological Conformal
 Symmetry in Non-critical $W_3$ Strings, UTHEP-277, May 1994,
 hep-th/9405049.

\end{thebibliography}
\end{document}